# The Multidimensional Assessment of Scholarly Research Impact

Henk F. Moed, Informetric Research Group, Elsevier | Radarweg 29, 1043 NX Amsterdam, The Netherlands | Email: h.moed@elsevier.com | Phone: +31 20 485 3436

Corresponding author:
Gali Halevi, Informetric Research Group, Elsevier | 360 Park Av. South, New York NY 10011| Email: g.halevi@elsevier.com | Phone: +1 646 248 9464

## Abstract

This article introduces the Multidimensional Research Assessment Matrix of scientific output. Its base notion holds that the choice of metrics to be applied in a research assessment process depends upon the unit of assessment, the research dimension to be assessed, and the purposes and policy context of the assessment. An indicator may by highly useful within one assessment process, but less so in another. For instance, publication counts are useful tools to help discriminating between those staff members who are research active, and those who are not, but are of little value if active scientists are to be compared one another according to their research performance. This paper gives a systematic account of the potential usefulness and limitations of a set of 10 important metrics including altmetrics, applied at the level of individual articles, individual researchers, research groups and institutions. It presents a typology of research impact dimensions, and indicates which metrics are the most appropriate to measure each dimension. It introduces the concept of a "meta-analysis" of the units under assessment in which metrics are not used as tools to evaluate individual units, but to reach policy inferences regarding the objectives and general set-up of an assessment process.

## Section 1: Introduction

*The increasing importance of research assessment*

In the current economical atmosphere where budgets are strained and funding is difficult to secure, ongoing, diverse and wholesome assessment is of immense importance for the progression of scientific and research programs and institutions. Research assessment is an integral part of any scientific activity. It is an ongoing process aimed at improving the quality of scientific-scholarly research. It includes evaluation of research quality and measurements of research inputs, outputs and impacts, and embraces both qualitative and quantitative methodologies, including the application of bibliometric indicators and peer review. Assessment and evaluation methods are widely used in the academic arena as they relate to student performance. However as Ewell (2009) pointed to, there is a distinction between external and internal assessments and their goals which is valid in this content as well. Assessment for *improvement* is an internal matter, aimed to gain insight into how well students participate in programs and use the knowledge gained. The results of such assessment are then used to improve the pedagogical approaches to learning accordingly. In contrast, assessment for purposes of *accountability* is used primarily to demonstrate that the institution is using its resources appropriately to help students develop the knowledge and function effectively in the 21st century (Ewell, 2009, p. 4-5).

This distinction is a valid also in the context of multi-dimensional assessment approach presented in this paper and is a part of the considerations of methods and data that should be applied according to the purpose of evaluation. In the framework presented here, assessment is seen as a combination of methods that can be modeled, changed and combined in different manners over time to insure that the final evaluation of a scientific program, department or even individual researcher is a positive one and in line with the goals and objectives laid out for them. Research is considered a key factor in securing positive economic and societal effect of science (Miller & O'Leary, 2007). The manner by which research assessment is performed and executed is a key matter for a wide range of stakeholders including program directors, research administrators, policy makers and heads of scientific institutions as well as individual researchers looking for tenure, promotion or to secure funding to name a few. These stakeholders have also an increasing concern regarding the quality of research performed especially in light of competition for talent and budgets and mandated transparency and accountability demanded by overseeing bodies (Hicks, 2009).

There are several trends that can be identified in this context:

<u>Performance-based funding</u>: funding of scientific research – especially in universities – tends to be based more frequently upon performance criteria, especially in countries in which research funds were in the past mainly allocated to universities by the Ministry responsible for research as a bl*o*ck grant, the amount of which was mainly determined by the number of enrolled students. It must be noted that in the U.S. there has never been a system of block grants for research; in this country research funding; was, and is still, primarily based on peer review of the content of proposals submitted to funding organizations.

Government agencies as well and funding bodies rely on evaluation scores to allocate research budgets to institutions and individuals. Such policy requires the organization of large scale research assessment exercises (OECD, 2010; Hicks, 2012) especially in terms of monetary costs, data purchasing, experts' recruitment and processing systems.

<u>Research in a global market</u>: research institutions and universities operate in a global market. International comparisons or rankings of institutions are being published on a regular basis with the aim to inform students, researchers and knowledge seeking external groups. Research managers use this information to benchmark their own institutions against their competitors (Hazelkorn, 2011).

<u>Internal research assessment systems</u>: More and more institutions implement internal research assessment processes and build research information systems (see for instance EUROCRIS, 2013) containing a variety of relevant input and output data on the research activities within an institution, enabling managers distribute funds based on performance.

<u>Usage based assessments via publishers' sites</u>: Major publishers make their content electronically available on-line, and researchers as well as administrators are able to measure the use of their scientific output as a part of an assessment process (Luther, 2002; Bo-Christer & Paetau, 2012).

Construction of large publications repositories: Disciplinary or institutionally oriented publication repositories are being built, including meta-data and/or full text data of publications made by an international research community in a particular subject field, or by researchers active in a particular institution, respectively (Fralinger & Bull, 2013; Burns, Lana & Budd,2013).

Scientific literature databases availability: While the Science Citation Index founded by Eugene Garfield (1964) and published by the Institute for Scientific Information (currently Thomson-Reuters' Web of Science) has for many years be the only database with a comprehensive coverage of peer reviewed journals, in the past years two more have been added among which are Elsevier's Scopus and Google Scholar.

Social media: More and more researchers use social media such as Twitter, Facebook, and blogs, to promote their work, communicate with each other and with the international scientific community. A series of indicators is being developed and made available often via small information companies (Chamberlain, 2013).

More indicators are becoming available: bibliographical databases implement bibliometric features such as author h-indexes; publication and citation counts based on data from the large, multi-disciplinary citation indexes. In addition to those, indicators based on the number of full text downloads are also available today. Furthermore, specialized institutes carry out indicator-based studies, some academic, some government and others in the private domain.

Desktop bibliometrics: Calculation and interpretation of science metrics are not conducted necessarily by bibliometric experts. "Desktop bibliometrics", a term coined by Katz and Hicks in 1997 (Katz & Hicks, 1997) and referring to an evaluation practice using bibliometric data in a quick, unreliable manner is becoming a reality.

*Scope, objectives and structure of this paper*

In an important article published in 1996, Ben Martin underlined the multi-dimensionality of basic research in terms of its nature and output (Martin, 1996). He focused on the scientific dimension, defined as the contribution scientists make to the stock of scientific knowledge; the principal indicators he used are publication, citation counts, and peer- ratings. He presented an assessment methodology of "converging partial indicators"*,* based on a model presented in an earlier paper (Martin & Irvine, 1983). *".... All quantitative measures of research are, at best, only partial indicators - indicators influenced partly by the magnitude of the contribution to scientific progress and partly by other factors. Nevertheless, selective and careful use of such indicators is surely better than none at all. Furthermore, the most fruitful approach is likely to involve the combined use of multiple indicators. However, because each is influenced by a number of 'other factors', one needs to try and control for those by matching the groups to be compared and assessed as closely as one can"* (Martin, 1996, p. 351).

In 2010 the Expert Group on the Assessment of University-Based Research, installed by the European Commission, published a report introducing the concept of a multi-dimensional research assessment matrix, built upon the notion of multi-dimensionality of research expressed in the above mentioned article by Ben Martin. But rather than focusing on one single output dimension and underlining the need to obtain convergence among a set of different indicators in order to produce valid and useful outcomes, the report aimed at proposing *"a consolidated multidimensional methodological approach addressing the various user needs, interests and purposes, and identifying data and indicator requirements"* (AUBR, 2010, p. 10). It is based on the notion that *"indicators designed to meet a particular objective or inform one target group may not be adequate for other purposes or target groups"*. Diverse institutional missions, and different policy environments and objectives require different assessment processes and indicators. In addition, the range of people and organizations requiring information about university based research is growing. Each group has specific but also overlapping requirements (AUBR, 2010, p. 51).

The aim of the current article is to further develop the notion of the multi-dimensional research assessment matrix, in the following ways:
(a) It gives a systematic account of the potential usefulness and limitations of metrics generally applied at the level of individual articles, researchers, research groups, departments, networks and institutions.
(b) It gives special attention to a set of 10 frequently used metrics in research assessment but also takes into account relatively new indicators such as the H-Index, 'usage' indicators

      based on full text download of articles from publication archives, and mentions in social media, often denoted as 'altmetrics'.

(c) Furthermore, it presents an extended typology of dimensions of research impact, and indicates which metrics are the most appropriate to measure a dimension. In this way it further expands on those parts of the matrix in the AUBR report that relate to the use of research impact indicators. It focuses on the purpose, objectives and policy context of research assessments, and demonstrates how these characteristics determine the methodology and metrics to be applied. For instance, publication counts are useful instruments to help discriminating between those staff members who are research active, and those who are not, but are of little value if research active scientists are to be compared one with another according to their research performance.

(d) The paper also introduces the concept of a "meta-analysis" of the units under assessment in which metrics are not used as tools to evaluate individual units, but to reach policy decisions regarding the overall objective and general set-up of an assessment process.

The principle assumption underlying this article is that the future of research assessment exercises lies in the intelligent combination of metrics and peer review. A necessary condition is a thorough awareness of the potentialities and limitations of each of each methodology.

This paper deals with basic research, primarily intended to increase scholarly knowledge, but often motivated by and funded for specific technological objectives such as the development of new technologies such as medical breakthroughs. Following Salter and Martin (2001), it includes both 'curiosity-driven' – sometimes also denoted as 'pure' – as well as 'strategic' or 'application oriented' research. The latter type of research may be fundamental in nature, but is undertaken in a quest for a particular application, even though its precise details are not yet known.

Metrics are standards of measurement by which efficiency, performance, progress, or quality of a plan, process, or product can be assessed (BusinessDictionary.com). In this article metrics (also denoted synonymously as indicators throughout this paper) are conceived as instruments used to measure the various components of research activity including inputs, process, outputs, efficiency, and impact and benefits (AUBR, 2020, p. 36).

The following are definitions of the various indicators used in measuring research activity:

*Input:* indicators that measure the human, physical and financial commitments devoted to research. Typical examples are the number of (academic) staff employed or revenues such as competitive, project funding for research.

*Process:* indicators that measure how research is conducted, including its management and evaluation. A typical example is the total of human resources employed by university departments, offices or affiliated agencies to support and fulfill technology transfer activities.

*Output:* indicators that measure the quantity of research products. Typical output forms are listed in Table 1.

*Research efficiency or productivity:* indicators that relate research output to input. Typical examples of metrics are the number of published articles per FTE research, or the number of citations per Euro spent on research.

*Impact:* refers to the contribution of research outcomes to the advancement of scientific-scholarly knowledge and to the benefits for society, culture, the environment or the economy.

This article focuses on the impact of research and a distinction is made between two main types of impact: scientific-scholarly and societal. The term 'societal' embraces a wide spectrum of aspects outside the domain of science and scholarship itself, including technological, social, economic, educational and cultural aspects. The various impact dimensions are further discussed in section 4.

| **Impact** | **Publication/text** | **Non-publication** |
|---|---|---|
| Scientific-scholarly | Scientific journal paper; book chapter; scholarly monograph | Research data file; video of experiment |
| Educational | Teaching course book; syllabus; text- or hand book | Skilled researchers (e.g., doctorates) |
| Economic or technological | Patent; commissioned research report; | Product; process; device; design; image; spin off |
| Social or cultural | Professional guidelines; newspaper article; communication submitted to social media, including blogs, tweets. | Interviews; events; performances; exhibits; scientific advisory work; |

Table 1: Main Types of Academic Research Output Forms.

Table 1 aims to present at least the most important output forms, and gives a typical examples of these, but it is far from being complete. Documents related to Research Excellence Framework (REF) in the UK give a more comprehensive overview of the output forms taken into account in the assessment of research in the various major disciplines (REF, 2012).

The structure of this article is as follows; section 2 presents the main types of metrics and analytical tools, and section 3 the main assessment models used in quantitative research assessment. Section 4 further develops the concept of the multi-dimensional research assessment matrix and Section 5 draws the main conclusions, and makes suggestions for future research.

## Section 2: Main types of metrics and analytical tools

*Introduction*

The assessment of scientific merit including individuals, institutions and program, departments, research groups and networks has a long a respectful history which was demonstrated in numerous methods and models utilizing different data sources and approaches (Abbott, et. al. 2010; Rohn, 2012; Vale, 2012; Zare, 2012). The proliferation and increasing amount and availability of primary data have created the ability to evaluate research on many levels and degrees of complexity on the one hand but also introduced some fundamental challenges to all who are involved in the process including evaluators, administrators and researchers on the other (Ball, 2007; Simons, 2008). Evaluative methods are used on several levels within the scientific world: (1) Institutional level (2) Program level and (3) Individual level. Each one of these levels has its objectives and goals. Institutional evaluation is being used in order to establish accreditation, define missions, establish new programs and monitor the quality of its research activities among others. The types of evaluative results can be seen in the ranking systems of universities which at present produced both regional and international indexes based on different criteria (Bornmann, et al., 2013; Lin, et. al., 2013; O'Connell, 2013; Pusser & Marginson, 2013).

Institutional evaluations are performed based on prestige measures derived from publications, citations, patents, collaborations and levels of expertise of the individuals within the institution. Program level evaluations are performed in order to measure the cost-benefit aspects of specific scientific programs. These are usually based on discovering the linkage between the investment made and the potential results of the program (Imbens & Wooldridge, 2008). Within this realm we find measures developed for technology transfer capabilities and commercialization potentialities of the program among others (Arthur & Blitz, 2000; Simpson, 2002; Rowan-Szal, et.al, 2007; Lane, 2010).

Finally an individual evaluation is mainly performed for purposes of promotion and retention of individuals, conducted in specific times in a researcher's career. Individual assessment methods rely on counts of publications or citations (Lee, et. al, 2012) as well as expert opinions which are a part of a peer-review process. In the past few years with the advent of social media, measures based on mentions in social media sites such as blogs, websites, Twitter and others which are labelled as "altmetrics" are on the rise (Galligan & Dyas-Correia, 2013; Osterrieder, 2013; Wang, et. al., 2013; Wilson, 2013).

*Publication- and citation-based indicators*

Methods involving counts of publications and citations are well known to the scientific community. These were and probably still are the main component of an institution or individual research assessment results. Citation analysis is one of the most established methods of evaluation (Van Raan, 1996; 2004). This type of analysis can measure the impact and intellectual influence of scientific and scholarly activities including: (1) publication impact; (2) author impact; (3) institution or department impact; and (4) country impact.

Using large sets of citations data can provide a reliable understanding of the intellectual influence of scientific output since when analyzing large sets of citations, random errors and individual variants are balanced. The authors of this paper agree with Zuckerman (1987) that, on the one hand, the presence of error does not preclude the possibility of precise measurement and that the net effect of certain sorts of error can be measured, but that on the other hand the crucial issue is whether errors are randomly distributed among all subgroups of scientists, or whether they systematically affect certain subgroups (Zuckerman, 1987, p. 331). Thus, it cannot a priori be assumed that any deviations of the norm cancel out when data samples are sufficiently large. For instance, Waltman, Van Eck and Wouters (2013) found systematic biases at the level of individual authors.

As with any type of statistical method, the use of citations analysis as an evaluative method should be used with caution mainly because of their biased nature. The list of references included in a scholarly work does not necessarily showcase all the literature that was read beforehand but not cited in the final product. There are also times when theoretical or well established works, despite providing the basis for publications are not cited by the author in what was labeled by McCain (2011) as "obliteration by incorporation". Other biases may include author self-citations, citations of a closed community or consensus among others (Bonzi & Snyder, 1991).

Large differences exist in publication and citation practices between subject fields. For instance, in molecular biology cited reference lists in scientific publications tend to be much longer than in mathematics, and more focused on recently publishes articles. As a result, citation rates of target articles in the former tend to be much higher than in the latter, especially during the first years after publication date. Absolute counts tend to be distorted by such differences, whereas normalised indicators can properly take them into account. A typical example of a normalised citation impact indicator is one that relates a group's citation impact to the world citation average in the subfields in which it is active. (e.g., Moed, 2005, p. 74). Other approaches based on citation percentiles calculated by discipline can for instance be found in Bornmann and Marx (2014).

Publications and citations analysis could have been a straight forward method, but years of bibliometric research have proven that it is not the case at all. Name variations of institutions and individuals make it difficult to correctly count these. Limited coverage or lacking coverage of the database selected for the analysis can cause fundamental errors. In addition, documents such as technical reports or professional papers which some label as "grey literature" are usually excluded from the analysis due to lack of indexing which may cause, in certain disciplines, a partial assessment. During the past 10 years new citation metrics were introduced, including the H-Index (Hirsch, 2005), G-Index and the i10-Index (Google, 2011) to name just a few.

Furthermore, the prestige of the journal in which a paper is published introduces yet another complication. There are several indexes which are used to establish the prestige of journals, among which are the Thomson Reuters Journal Impact Factor (Garfield, 1994), SNIP (Moed, 2010), and SJR (Gonzalez-Pereira, Guerrero-Bote, & Moya-Anegon, 2009). Each of these methods addresses different challenges that such assessment produces. The debate about the fairness and accuracy of citations indices as measures of productivity and impact will probably not subside soon and despite the opposition and obvious limitations of such measures, counts are still wildly used as research assessment method.

*Usage-based indicators*

Usage data is also becoming available for analysis. In their 2010 review article, "Usage Bibliometrics", Kurtz and Bollen explored the addition of modern usage data to the traditional data used in bibliometrics e.g. publications data. The usage data includes clickstreams, downloads and views of scholarly publications recorded on an article level. A well-known phenomenon in research is the difference between the amount of articles read, browsed or scanned and the number of times it is cited. This can change from discipline to discipline and from one institution to another depending on its reading and citing behaviour. In addition, there are certain types of documents that might be read more than cited such as reviews, editorials, tutorials or other technical output (Paiva, et.al, 2012, Schloegl & Gorraiz, 2010).

The fact that these articles may not be cited as often as their counterparts cannot be indicative of impact, as obviously they reach a large audience. Usage data can, in principle point to such works and allow its authors to be recognized for their contribution.

Citations, publications and usage are seen to be combined and calculated in order to develop models that can capture the weight of each one and provide better understanding on the relationship between them and how those can be applied to an institution and individual's assessments (Bollen & Van De Sompel, 2008). There are several challenges to usage analysis among which are: (1) incomplete data availability across providers; (2) differences between disciplines and institution reading behaviours which are difficult to account for; (3) content crawling and automated downloads software tools that allow individuals to automatically crawl and download large amount of content which doesn't necessarily mean that it was read or viewed; and (4) the difficulty to ascertain whether downloaded publications were actually read or used. Hence, analyzing combined sets of usage and citations is gaining ground in bibliometric research as a way to provide an accurate dimension to the evaluation process.

*Altmetrics*

Altmetrics is a relatively new area in metrics development. It emerged from the increasing numbers of social media platforms and their prolific use by scientists and researchers (Taylor, 2013).One type of altmetrics counts the number of times a publication is mentioned in blogs, tweets or other social media platform such as shared references management websites. Different weights are assigned to each data source and altmetric scores are given to individual publication. The rationale behind the use of these alternative measures is that mentions of a publication in social media sites can be counted as citations and should be taken into consideration when reviewing the impact of research, individual or institution (Adie & Roe,2013; Barjak, et.al, 2007). Today there are a few companies that offer altmetric score calculation such as Altmetric.com and ImpactStory.com. Implementation of altmetrics indicators in scientific databases is increasing and examples can be seen in Scopus.com, PLOS and more. There are several types of data used to measure the way scholarly publications are used (blog.impactstory.com):

1. Views - HTML views and PDF downloads
2. Discussion - journal comments, science blogs, Wikipedia, Twitter, Facebook and other social media
3. Saving & Sharing - Mendeley, CiteULike and other social bookmarks
4. Recommendation - for example used by Faculty of 1000 (F1000)

*Patent-based indicators*

Patent analysis is a unique method that not only measures the number of patents associated with an institution or an individual but also looks at its citations from two perspectives: (1) the non-literature citations; and (2) patent to patent citations (Narin, 1994). Modern technology is more and more science-based, and academic researchers increasingly appear as inventors of patents (Schmoch, 2004). Similarly to citation analysis of journal articles, patent analysis identifies high citations to basic and applied research papers in patents as well as patents that are highly cited by recently issued patents. This method attempts to provide a direct linkage between basic and applied science and patents as indication of economic, social and/or methodological contribution (Narin & Olivastro, 1997; Michel & Bettels, 2001).

The challenges associated with this method are quite similar to those found in the citation analysis of journal articles. Incomplete and un-standardized data, patenting in countries other than where the institution or individual originates from and lack of exhaustive reference lists within the patents are just a few. These are of course major limitations that may have negative effects on the institution or individual assessments. Yet, patents are almost the only form of public communication that can be used as indicator of technological innovation and thus it is used as a part of the evaluation of institutions and individuals.

*Network-based indicators*

Networks analysis is one of the more recent method used for scientific assessment. The technological capability to trace and calculate collaborations between institutions and individuals on a large scale and through the years enables evaluators to have a novel view on how institutions and individuals work as a part of the domestic and global research network (Bozeman, Dietz & Gaughan, 2001; Martinez, et.al, 2003). It is assumed that institutions and individuals who develop and maintain a prolific research network are not only more productive but also more active, visible and established.

The network analysis also allows benchmarking to be performed by evaluators by comparing collaborating individuals and institutions to each other. This type of comparison puts their research output and impact in context of the domestic and international disciplinary activity and allows for better understanding of their rank among their peers. Such network analytics is done on publication level but also, now, on social media and public domain level where the scientific community shares outcomes and accomplishments openly.

Research mobility is also among the network analytics methods (Zellner, 2003; Ackers, 2005). This method enables tracing an individual's affiliations through the years and look at his/her expertise building throughout a career. It is assumed that moving from one institution to another throughout different stages of one's career helps in expertise building and can result in high productivity. Of course, there are many challenges to this approach and its value is still examined. One of the main challenges resides in the fact that affiliations' names as mentioned in the publish papers are not always standardized, thus making it difficult to trace. Another factor is that education in a different country which might not have resulted in a publication cannot be measured, thus making this particular expertise building impossible to trace.

*Economic indicators*

Econometric tools and models aim to measure the effect of science on industry, innovation and the economy as a whole. Within this evaluative framework one finds on the metrics side technology transfer measures and patentability potentialities of a research carried as well as cost-benefit measures. The global economic crisis of the past decade has brought this type of evaluation to the front so that programs and institutions are not only evaluated on the basis of their contribution to science but also on the level of their contribution to the industry and commerce.

Commercialization of research via patents, start-up companies formation has been a topic of research and analysis for quite some time (Chen, Roco & Son, 2013; Huang et al. 2013), however, the use of these measures are now being brought forth into the evaluation arena because of two reasons: (1) the availability of and ability to collect and analyse large scale datasets including patents, financial and technical reports globally; (2) the increasing demand by the public and government to demonstrate cost-benefit measures of programs within scientific institutions especially those publically funded.

Measures such as these are not without their challenges. First, the statistical models used are complex and require deep understanding of the investment made but also of the program itself. Long term programs are more difficult to measure as far as the cost-benefit or even tech-transfer is concerned and requires expertise not only in mathematics and statistics but also in the field of investigation itself.

*Big data and its effect on evaluation methods*

Big data refers to is a collection of data sets that are so large and complex that it becomes difficult to process using on-hand database management tools or traditional data processing applications. The advent of super computers and cloud computing able to process, analyze and visualize these datasets has its effect also on evaluation methods and models. While a decade ago, scientific evaluation relied mainly on citations and publications counts, most of which were even done manually, today this data is not only available digitally but can also be triangulated with other data types (Moed, 2012). Table 2 depicts some examples of big datasets that can be combined in a bibliometric study to investigate different phenomena related to publications and scholarly output.

Thus, for example, publications and citations counts can be triangulated with collaborative indicators, text analysis and econometric measures producing multi-level view of an institution, program or an individual. Yet, the availability and processing capabilities of these large dataset does not necessarily mean that evaluation becomes simple or easy to communicate. The fact of the matter is that as they become more complex, both administrators and evaluators find it difficult to reach consensus as to which model best depicts productivity and impact of scientific activities. These technological abilities are becoming breading ground to more indices, models and measures and while each may be valid and grounded in research they present a challenge in deciding which is best to use and in what setting. Table 2 lists some of the most frequently used data types for each of the above listed evaluation methods.

| Combined datasets | Studied phenomena | Typical research questions |
|---|---|---|
| Citation indexes and usage log files of full text publication archives | Downloads versus citations; distinct phases in the process of processing scientific information | What do downloads of full text articles measure? To what extent do downloads and citations correlate? |
| Citation indexes and patent databases | Linkages between science and technology (the science–technology interface) | What is the technological impact of a scientific research finding or field? |
| Citation indexes and scholarly book indexes | The role of books in scholarly communication; research productivity taking scholarly book output into account | How important are books in the various scientific disciplines, how do journals and books interrelate, and what are the most important books publishers? |
| Citation indexes (or publication databases) and OECD national statistics | Research input or capacity; evolution of the number of active researchers in a country and the phase of their career | How many researchers enter and/or move out of a national research system in a particular year? |
| Citation indexes and full text article databases | The context of citations; sentiment analysis of the scientific-scholarly literature | In what ways can one objectively characterize citation contexts? And identify implicit citations to documents or concepts? |

Table 2: Compound Big Datasets and their Objects of Study. Research Trends, Issue 30, September 2012
http://www.researchtrends.com/issue-30-september-2012/the-use-of-big-datasets-in-bibliometric-research/

*Costs of Research Evaluation*

A well-constructed and executed research evaluation process incurs significant costs regardless of the methodology used. Hicks (2012), reviews some of the costs related to the implementation and use of performance-based university research funding systems (PRFSs) which are national systems built to evaluate a country's scientific output in order to better allocate funding based on performance. Since it is difficult to estimate the cost of each research evaluation method (Hicks, 2012, p.256), taking the PRFS and their related costs, provides an understanding of the investment needed in order to conduct a sound scientific evolution as these systems include different data and analytical methodologies (Hicks, 2012, p. 255). Peer review PRFSs incur indirect costs which include large panels of experts' time needed for the compilation and review of a university's output and faculty time needed for preparing submissions. However, indicators-based systems also incur costs mainly due to the need of building, cleaning and maintaining s documentation system, purchasing citations data from vendors and developing calculation algorithms (Hicks, 2012, p. 258). This is true for any data-based evaluation method. Procuring the data, cleaning it and embedding it in a sound system are only part of the costs involved. Developing advanced algorithmic calculations of the data that will provide a true view of a country or an institution scientific output require expert opinion and know-how which come at a cost as well.

These expenses as well as locating and engaging with expert reviewers, resulted in what is referred to by Van Raan (2005) as "quickies", i.e. rapid, cheap evaluations based on basic documents and citations counts with the help of standard journal impact factors (Van Raan, 2005, p. 140). As Van Raan notes: "Quite often I am confronted with the situation that responsible science administrators in national governments and in institutions request the application of bibliometric indicators that are not advanced enough…the real problem is not the use of bibliometric indicators as such, but the application of less-developed bibliometric measures" (p.140-141).

Therefore, when considering an evaluative method and especially one that requires a combination of more than one methodology or data type, one has to carefully estimate and calculate the costs involved. From data purchasing to systems development to expert reviewers; all involved will require appropriate funding in order to avoid a 'quick and cheap' evaluation exercise that might hinder an institution's or individuals proper assessment.

## Section 3: Assessment models

*Base distinctions*

The AUBR Report makes two base distinctions regarding the types of assessment methodologies. The *first* relates to the assessment method itself, namely between peer review, providing a judgment based on expert knowledge, and a metrics-based assessment, using various types of indicators, including bibliometric, econometric, and altmetric measures, and also end-user reviews, measuring for instance customer satisfaction. A second distinction relates to the role of the research unit under assessment in the evaluation process. It differentiates between self-evaluation – defined as a form of self-reflection which involves critically reviewing the quality of one's own performance – and external evaluation, conducted by an external evaluation agency. Methods are often combined, and are then to be viewed as components of an integral research assessment methodology.

The challenges mentioned in section 2 above, including the evaluation method, data and analytics selection have brought forth the development of hybrid evaluation models. These models aim to build a modular method combining different measures and approaches depending on the field and target of assessment. These models were built for specific disciplines or areas of investigation answering challenges arising there. However, their modular nature and comprehensive approach demonstrate the importance of utilizing a variety of measures, models and method in order to accurately capture the impact and productivity of institutions, programs and individuals. Below are some examples of hybrid evaluation models.

*Program Assessment - Empowerment Evaluation (EE)*

EE is a flexible, goal-oriented approach to evaluation that puts an emphasis on the people involved. It places both evaluators and those being evaluated on the same levels of involvement and commitment to the success of individuals, programs and institutions. This method was conceived in 1992 (Fetterman, 1994) and is being continuously developed since. It can be applied to a variety of activities and performances, not merely to scientific research. EE works on several principles which mainly aim to have all involved (evaluators included) as stakeholders in the evaluation process. Table 3 summarizes the EE principles.

| Category | Principle |
|---|---|
| Core Values | EE aims to influence the quality of programs<br>Power and responsibility for evaluation lies with the program stakeholders<br>EE adheres to the evaluation standards |
| Improvement-oriented culture | Empowerment evaluators demystify evaluation<br>Empowerment evaluators emphasize collaboration with program stakeholders<br>Empowerment evaluators build stakeholders capacity to conduct evaluation and use results effectively<br>Empowerment evaluators use evaluation results in the spirit of continuous improvement |
| Developmental Process | EE is helpful in any stage of program development<br>EE influences program planning<br>EE institutionalizes self- evaluation |

Table 3: Summary of EE principles; Wandersman, A., et.al, 2004. Empowerment evaluation: Principles and action. *Participatory community research: Theories and methods in action. Washington, DC: American Psychological Association*. Page 141

The unique attributes of this model lays in its capability to combine social, humanistic and traditional evaluative approaches in a holistic manner. In addition, this model can be easily implemented in different areas; from scientific to social programs to industry performance and more (Wandersman & Snell-Johns, 2005; Fetterman &Wandersman, 2007). The stakeholders are responsible to the selection of methods and metrics appropriate to the purpose of assessment and take active part in not only collecting and analyzing related data but also understanding it and implementing improvements processes.

*Field-specific Evaluation: The Becker Model*

This model was developed by Cathy Sarli and Kristi Holmes at the Bernard Becker Medical Library at Washington University. It aims to provide a framework for tracking diffusion of research outputs and activities and identify indicators that demonstrate evidence of biomedical research impact. It is intended to be used as a supplement to publication analysis. The model consists of four dimensions within which a variety of indicators are utilized: (1) research output; (2) knowledge transfer; (3) clinical implementation; and (4) community benefit. Indicators that demonstrate research impact are grouped at the appropriate stages along with specific criteria that serve as evidence of research impact for each indicator. By using a multilevel approach to evaluating biomedical research impact the model aims to be able and assist scientists and program managers identify return on investment, quality of publications, collaborations opportunities to name a few (The Becker Model: https://becker.wustl.edu/impact-assessment/model)

*University Ranking Models*

There are many systems and models that rank universities' prestige compared to their peers. Most of these approaches are based on scientific output (bibliometric) measures and indicators. The "Academic Ranking of World Universities" (ARWU) is produced by the Centre of World-Class Universities at Shanghai Jiao Tong University. For over a decade, ARWU has been presenting the world Top 500 universities worldwide (Liu & Cheng, 2005). Examples of academic and private research institutions' indicators development include the Leiden rankings, High Impact Universities: Research Performance Index and the SCImago Institutions Rankings (http://www.scimagoir.com/).

The CWTS Leiden ranking http://www.leidenranking.com/ measures the scientific performance of 500 major universities worldwide. The aim of the model is to measure the scientific impact of universities and takes into consideration their collaboration with the scientific community. Others look at the professional dimension (i.e. Professional Ranking of World Universities http://www.mines-paristech.fr/Ecole/Classements/ and Human Resources & Labor Review http://www.chasecareer.net/) and some at web impact (i.e. the G-factor and the Webometrics Ranking of World Universities http://www.webometrics.info/about_rank.html). There are also several such evaluative models that take a multilevel approach to evaluation and calculate it based on various dimensions including social and economic impact. For example, The World University Rankings (http://www.timeshighereducation.co.uk/world-university-rankings/) uses a variety of indicators to determine the quality and prestige of higher education institutions.

The Global Research Benchmarking System www.researchbenchmarking.org (GRBS) developed by the International Institute for Software Technology (IIST) at the United Nations University in Macau. *The aim of this system is to "provide objective data and analyses to benchmark research performance in traditional disciplinary subject areas and in* *interdisciplinary areas for the purpose of strengthening the quality and impact of research".* This system does not give a ranking of universities but rather an information system which provides a tool for institutions to use in order to compare themselves to others in their research activity. The U-Multirank (http://www.u-multirank.eu) project is another example of a methodology that takes into account more than one aspect of an institution's scientific activities to arrive at a ranking. Some of the components taken into calculations are teaching and learning, research, knowledge transfer, international orientation and regional engagement. This approach is participant-driven and does not calculate indicators but rather compares similar institutions to each other on several levels. At the moment this project is sponsored by the European commission and includes 500 institutions from around the world.

These rankings and their findings are debatable and pro and cons regarding them are being discussed by both the scientific community as well as administrators (Van Raan, 2005; Calero-Medina, López-Illescas, Visser & Moed (2008). Billaut, Bouyssou & Vincke, 2010). However, the fact that more and more global rankings are relaying on a diverse set of indicators and measures demonstrates the overall agreement that no one indicator can capture quality or impact accurately.

## Section 4: The multi-dimensional research assessment matrix

*Base principles*

When building a research assessment process, one has to decide which methodology should be used, which indicators to calculate, and which data to collect. Therefore, one should address the following key questions as their answers determine which methodology and types of indicators should be used. Each question relates to a particular aspect of the research assessment process.

- What is the unit of the assessment? A country, an institution, a research group, an individual, or a research field or an international network? In which discipline(s) is it active?
- Which dimension of the research process must be assessed? Scientific-scholarly impact? Social benefit? Multi-disciplinarity? Participation in international networks?
- What are the purpose and the objectives of the assessment? Allocate funding? Improve performance? Increase regional engagement? Which "meta assumptions" can be made on the state of the units of assessment?

The goals set out to be achieved by the evaluating body should direct the process by which the assessment procedure is set out. Taking that into account, the evaluative body must take into consideration the principles offered here which are that the unit of assessment, the research dimension to be assessed, and the purposes of the assessment jointly determine the type of indicators to be used. An indicator may by highly useful within one assessment process, but less so in another. The aim of this section is to further develop this principle by taking into account new bibliometric and non-bibliometric indicators, a series of aggregation levels, impact sub-dimensions, and by focusing on the objectives and policy background of the assessment.

Two characteristics of the unit under assessment must be underlined, as they determine the type of measures to be used in the assessment. Firstly, the *discipline(s)* in which the unit under evaluation must be taken into consideration. There are several disciplines which are difficult to assess mainly because they are geographically or culturally specific. Among these one can identify linguistics (Nederhof, Luwel and Moed, 2001), language-specific literature, law et al., and others, especially in the humanities (Moed, Nederhof and Luwel, 2002). Secondly, the *mission* of the research unit under assessment is relevant as well. To the extent that it is taken into account in the assessment process, it determines the indicators that have to be applied.

*Potential usefulness and limitations of 10 frequently used indicators*

Table 4 summarizes the description of main types of indicators outlined in section 2, and gives some of the strong points and limitations of 7 publication- and citation-based indicators, a patent-based indicator and two altmetrics. More details can be found in section 2.

| *Indicator* | *Potentialities; strong points* | *Limitations* |
|---|---|---|
| Number of published articles | This is a useful tool to identify lagging research units if the metric's value is below a certain (subject field dependent) minimum | If numbers exceed a certain minimum level, differences between them cannot be interpreted in terms of performance |
| Number of citations | Useful for weighting individual publications. Reveals impact of the total collection of a research group's articles, disregarding how citations are distributed among cited articles | Depends upon subject field and age of (cited) publications. Depends upon the size of the group's publication volume |
| Citations per article | Reveals influence relative to size of publication volume | Strongly depends upon subject field and age of cited articles, and also upon type of document (e.g., normal article versus review). |
| Normalized citation rate | Takes into account type (e.g., review, full length article), subject field and age of cited article | Field delimitation must be sound. Should be used with special caution when comparing units with very different publication volumes or active in highly specialized subjects |
| Indicators based on Citation percentiles- (e.g., top 10 % ) | Focuses on the most important publications; does not use the mean of (skewed) citation distributions; normalizes outliers | Maps all actual values onto a 0-100 scale; one may lose the sense of underlying absolute differences, and undervalue extraordinary cases |
| Journal impact factor and other journal metrics | The quality or impact of the journals in which a unit has published is a performance aspect in its own right | Journal metrics cannot be used as a surrogate of actual citation impact; impact factors are no predictors of the citation rate of individual papers |
| H-Index | Combines an assessment of both quantity (nr. papers) and impact (citations). Tends to be insensitive to highly cited outliers and to unimportant (uncited) articles | Its value is biased in favor of senior researchers compared to juniors; actual impact of the most cited papers hardly affects its value |
| Number of patents | Inventions may be disclosed in patents; patent data is available at a global level | Not all inventions are patentable or actually patented. The number of patents filed differs across countries because of legislation or culture, and also across subject fields |
| Full text article download counts | Are available almost immediately after publication; may reveal use or value that is not expressed in citations, impact upon scholarly audiences from other research domains or upon non-scholarly audiences | Downloaded articles may be selected according to their face value rather than their value perceived after reflection; |
| Mentions in social media | Are immediately available after publication; may reveal impact upon scholarly audiences from other research domains or upon non-scholarly audiences | Scientific-scholarly and societal impact are distinct concepts. One cannot measure scientific-scholarly impact with metrics based on social media mentions. |

Table 4: Potentialities and Limitations of 8 Frequently Used Bibliometric and 2 Altmetrics Indicators.

*Units of assessment and the role of metrics in general*

Table 5 presents the potentialities and limitations of the use of metrics for five units of assessment at different aggregation levels. Most limitations relate to the network structure among units of assessment, and underline that a particular unit must be viewed within the context of the network in which it takes a part. For instance, individual research papers are not isolated entities, but can be viewed as elements of publication oeuvres of research groups; citations to a single key paper may aim to acknowledge the total oeuvre (Moed, 2005). Researchers tend to operate in teams and therefore an assessment of their individual performance should take this into account. Non-bibliometric indicators may be used as a way to reflect more personal achievements, such as invitations for lectures at international conferences or at seminars in prestigious institutions. Universities in countries with a strong research infrastructure outside the university system, tend to gain less visibility in international university rankings than universities in countries in which research is mainly concentrated in the academic sector.

| *Unit of Assessment* | *Metrics Potentialities* | *Metrics Limitations* |
|---|---|---|
| Individual article | Metrics reveal differences in significance between articles and may identify key articles | Individual articles are not isolated entities but rather elements of publication oeuvres; different types of articles exist. |
| Individual author | Metrics reveal differences in impact between individuals | Most research articles are the result of team work and are multi-authored. How do we then assess the role of an individual in a team? |
| Research group | The research group is the core research entity, at least in science | Social sciences and humanities do not always show a group structure as in science |
| Research Institution | Metrics show status and impact of research institutions | Institutions may specialize or be more general, and have specific functions in a national research system; large differences may exist within institutions |
| Country | Metrics unravel the structure of national research systems | Aggregate data may conceal differences between a country's research institutions |

Table 5: Main Units of Assessment and the Role of Metrics.

*Research dimensions and its principal indicators*

The variety of impact dimensions is presented in Table 6 which distinguishes the various types of research impact, and gives typical examples of indicators that may be used to assess these. The two main categories are scientific-scholarly and societal impact. The term 'societal' embraces a wide spectrum of aspects outside the domain of science and scholarship itself, including technological, social, economic, environmental, and cultural aspects. The list of indicators includes the 10 metrics that are given special attention in this paper, and also a number of other indicators, partly derived from the AUBR Report, but it do not claim to be fully comprehensive.

| Type of impact | Short Description; Typical examples | Indicators (examples) |
|---|---|---|
| **Scientific-scholarly or academic** | | |
| Knowledge growth | Contribution to scientific-scholarly progress: creation of new scientific knowledge | Indicators based on publications and citations in peer-reviewed journals and books |
| Research networks | Integration in (inter)national scientific-scholarly networks and research teams | (inter)national collaborations including co-authorships; participation in emerging topics |
| Publication outlets | Effectiveness of publication strategies; visibility and quality of used publication outlets | Journal impact factors and other journal metrics; diversity of used outlets; |
| **Societal** | | |
| Social | Stimulating new approaches to social issues; informing public debate and improve policy-making; informing practitioners and improving professional practices; providing external users with useful knowledge; Improving people's health and quality of life; Improvements in environment and lifestyle; | <ul><li>Citations in medical guidelines or policy documents to research articles</li><li>Funding received from end-users</li><li>End-user esteem (e.g., appointments in (inter)national organizations, advisory committees)</li><li>Juried selection of artworks for exhibitions</li><li>Mentions of research work in social media</li></ul> |
| Technological | Creation of new technologies (products and services) or enhancement of existing ones based on scientific research | Citations in patents to the scientific literature (journal articles) |
| Economic | Improved productivity; adding to economic growth and wealth creation; enhancing the skills base; increased innovation capability and global competitiveness; uptake of recycling techniques; | <ul><li>Revenues created from the commercialization of research generated intellectual property (IP)</li><li>Number patents, licenses, spin-offs</li><li>Number of PhD and equivalent research doctorates</li><li>Employability of PhD graduates</li></ul> |
| Cultural | Supporting greater understanding of where we have come from, and who and what we are; bringing new ideas and new modes of experience to the nation. | <ul><li>Media (e.g. TV) performances</li><li>Essays on scientific achievements in newspapers and weeklies</li><li>Mentions of research work in social media</li></ul> |

Table 6: Types of Research Impact and Indicators

*Assessment purpose and objectives; the role of the policy context*

In this section it is argued that the selection of the indicators in a research assessment exercise very much depends upon the policy context in which the assessment takes place. In addition, it depends on the "state" or "condition" of the unit(s) of assessment. This claim is illustrated below with two examples relating to the assessment of individuals: one relates to the use of journal metrics, and a second to the application of publication counts. The policy relevance of these examples is that managers at the departmental and central level in academic institutions are confronted with the necessity to evaluate researchers for promotion or hiring on a daily basis.

A distinction can be made between *purpose* and *objective* of an assessment. A purpose has a more general nature, and tends to be grounded in general notions (e.g., "increase research performance"), whereas objectives are more specific, more formulated in operational terms (e.g., "stimulate international publishing"). Objectives are grounded in assumptions on how they are relate to the general purpose (e.g., "it is assumed that by stimulating international publishing, research performance increases, at least at the longer run").

The policy relevance of both assessment purposes and objectives follows from what may be termed as a "meta assumption" on the state of the units of assessment, which in turn, is based on a Meta-analysis of these units. For instance, "stimulating international publishing" as an objective in a national research assessment exercise makes sense from a policy viewpoint only if there are good reasons to believe that the level of international publishing among a country's researchers is relatively low compared to their international counterparts. Similarly, assessing whether an academic staff member is "research active" or not, seems to make sense only of there is evidence that a non-negligible part of staff hardly carries out research.

"International publishing" may relate to the level of the quality criteria applied by editors and referees in the review of submitted manuscripts, or to the geographical location of authors, members of the editorial or referee board, and/or readers of a journal. The following definition would include both dimensions: international publishing is publishing in outlets that have: (1) rigorous, high-standard manuscript peer review; and (2) international publishing and reading audiences.

Bibliometric studies found that the journal impact factor is a proxy of a journal's international status. For instance, Sugimuto et al. (2013) reported that acceptance rates of manuscripts submitted to scientific journals negatively correlate with the journals' impact factors, suggesting that journals with rigorous referee systems tend to generate higher impact than others.

If an analysis of the state of a country's science concludes that a substantial group of researchers tends to publish predominantly in national journals that are hardly read outside the country's borders and do not have severe rigorous peer review, it is in the view of the authors of this paper, defendable to use the number of publications in the top quartile of journals according to citation impact as an indicator of research performance. In this manner one is able to discriminate between those researchers whose research quality is sufficiently high to publish in international, peer reviewed journals, and those who are less capable of doing so. This issue is further discussed in Section 5. But if in internationally oriented, leading universities one has to assess candidates submitting their job application, it is questionable whether it makes sense comparing them according to the average citation impact of the journals in which they published their papers, using journal impact factors or other journal indicators. Due to self selection, the applicants will probably publish at least a large part of the papers in good, international journals. Other characteristics of the published articles, especially their actual citation impact, are probably more informative as to the candidates' past research performance and future potential than indicators based on journal metrics are.

A second example relates to the use of publication counts. In order to identify academic staff that is not research active, it is reasonable to consider the publication output of the staff under assessment, and identify those whose output is below a certain – subject field dependent – minimum. But if one has to assess candidates submitting their job application to a leading research university, it hardly makes sense to compare them according to their publication counts. Due to self-selection, they will probably all meet a minimum threshold. In other words, while there are good reasons to believe that journal metrics or publication counts are appropriate indicators to identify the bottom of the quality distribution of research staff, they have a limited value if one aims to discriminate in the top of that distribution.

These examples illustrate that the choice of indicators depends not only upon the overall purpose of the assessment, but also upon the specific objectives, and on the Meta view on the state of the units of assessment. These factors are best be characterized by the term "policy context". Therefore, the conclusion is that the selection of indicators in an assessment depends upon the unit of assessment, the research aspect to be assessed, and very much on its policy context.

## Section 5: Discussion and conclusions

*Meta-analysis*

It was stated that a meta-analysis of the "state of the units of assessment" determines the methodology and indicators to be applied in an assessment process. It must be noted that bibliometric indicators and other science metrics may – and actually do - play an important role in the empirical foundation of such a Meta view. Metrics are essential tools on *two* levels: in the assessment process itself, and on the Meta level aimed to shape that process. Yet, their function in these two levels is different. In the first they are tools in the assessment of a particular unit, e.g., a particular individual researcher, or department, and may provide one of the foundations of evaluative statements about such a unit. At the second level they provide insight into the functionality of a research system as a whole, and help draw general conclusions about its state assisting in drafting policy conclusions regarding the overall objective and general set-up of an assessment process.

A Meta level analysis can also provide a clue as to how peer review and quantitative approaches might be combined. For instance, the complexity of finding appropriate peers to assess all research groups in a broad science discipline in a national research assessment exercise may urge the organizers of that exercise to carry out a bibliometric study first and decide on the basis of its outcomes in which specialized fields or for which groups a thorough peer assessment seems necessary. One important element of the Meta-analysis is a systematic investigation of the effects of the assessment process, both the intended and the unintended ones.

*Statistical considerations*

The observation that the usefulness of journal impact factors and publications counts so strongly depends upon a meta view of the units to be assessed, can also be grounded in statistical considerations. If in a particular study a positive (linear or rank) correlation is found to hold between two variables, it does not follow that it holds for all sub-ranges of values of the variables. Whether or not a sample of the two variables can be expected to correlate in a particular study, very much depends upon the range of values obtained by the units in the sample.

For instance, Sugimoto et al. (2013) examined the relationship between journal manuscript acceptance rates and 5-year journal impact factors, and found in a sample of 1,325 journals a statistically significant linear correlation coefficient between these two measures. But, most importantly, the study also found that, when dividing journals into quartiles according to their acceptance rates and analyzing correlation coefficients *within* quartiles, the correlation coefficients between acceptance rates and impact factors were much lower and not significant. This shows that the application of journal metrics or publication counts to assess the comparative performance of researchers who publish on a regular basis in international journals cannot be sufficiently justified by referring merely to earlier studies reporting on observed positive correlation between these measures and peer ratings of research performance. What is not defendable in the view of the authors is the use of such indicators simply because they are relatively easy to calculate and readily available.

The authors of this paper share the critique, offered by The San Francisco Declaration on Research Assessment (DORA), for example, of the use of journal metrics in the assessment of individual researchers. Indeed, it does not make sense to discriminate in a group of research active researchers publishing in good journals between high and low performers on the basis of weighted impact factors of the journals in which they published their articles. On the other hand, it does not follow that the use of this type of indicator is invalid under all circumstances.

*Policy considerations*

Research assessments methodologies cannot be introduced in practice at any point in time, and do not have eternal lives. In the previous section it was stated that the authors of this article find it under certain conditions it defendable to use publication counts and journal metrics as one of the sources of information in individual assessments. But one may argue that it is fair to maintain a time delay of several years between the moment it is decided to use a particular assessment method or indicator on the one hand, and the time at which it is actually used, on the other. In this way, the researchers under assessment have the opportunity to change their publication behavior – to the extent that they are capable of doing that.

In recent years there have been several discussions that challenge the common practice of research evaluation using, for example, journal impact factors (Alberts, 2013; Van Noorden, 2013). The San Francisco Declaration on Research Assessment (DORA) is one of these manifestations, calling for improvements that need to be made to ways in which research is evaluated and especially challenging the impact factor as a tool in such evaluations. In the view of the authors of this paper it is wise to change an assessment method radically every 5 to 10 years. Two considerations may lead to such a decision. First, a meta-analysis may reveal that the overall state of the units of assessment has changed in such a manner, that the old methodology is either irrelevant or invalid. Secondly, any use of assessment methodologies and indicators must be thoroughly monitored in terms of its effects, especially the unintended ones. Severe negative effects such as manipulation of metrics may lead to the decision to abandon a method, and establish a new one, even though bibliometric can to some extent detect and correct for such behavior (Reedijk & Moed, 2008).

*What is an acceptable "error rate"?*

Regarding the – either negative or positive – effects of the use of metrics or any other methodology in research assessment, one may distinguish two points of view. One may focus on its consequences for an individual entity, such as an individual scholar, a research group or institution, or on the effects it has upon scholarly activity and progress in general. A methodology, even if it provides invalid outcomes in individual cases, may be beneficial to the scholarly system as a whole. Cole and Cole expressed this notion several decades ago in their study of chance and consensus in peer review of proposals submitted to the National science Foundation (Cole, Cole & Simon, 1981).

Each methodology has its strengths and limitations, and is associated with a certain risk of arriving at invalid outcomes. As Martin (1996) pointed out, this is true not only for metrics but also for peer review. It is the task of members from the scholarly community and the domain of research policy, and not of the authors to decide what are acceptable "error rates" and whether its benefits prevail, based on a notion of what is a fair assessment process.  Bibliometricians and other analysts of science and technology should provide insight into the uses and limits of the various types of metrics, in order to help scholars and policy makers to carry out such a delicate task.

## Acknowledgement

The authors wish to thank three anonymous referees for their valuable comments on an earlier version of this paper.